\documentclass[prb,twocolumn,floatfix,superscriptaddress]{revtex4}

\usepackage{graphicx}
\usepackage{color}
\usepackage[normalem]{ulem}
\usepackage{braket}
\usepackage{amsmath}
\usepackage{verbatim}
\usepackage{amssymb}

\newcommand{\be}{\begin{equation}}
\newcommand{\ee}{\end{equation}}
\newcommand{\ba}{\begin{eqnarray}}
\newcommand{\ea}{\end{eqnarray}}
\newcommand{\nn}{\nonumber \\}
\newcommand{\efb}{\Phi^x}
\newcommand{\ecjj}{\Phi_{\rm CJJ}^x}

\begin{document}

\title{Probing flux and charge noise with macroscopic resonant tunneling}

\author{Alexander M.~Whiticar}
\affiliation{D-Wave Systems Inc., 3033 Beta Avenue, Burnaby BC
Canada V5G 4M9}
\author{Anatoly Y.~Smirnov}
\affiliation{D-Wave Systems Inc., 3033 Beta Avenue, Burnaby BC
Canada V5G 4M9}
\author{Trevor Lanting}
\affiliation{D-Wave Systems Inc., 3033 Beta Avenue, Burnaby BC
Canada V5G 4M9}
\author{Jed Whittaker}
\affiliation{D-Wave Systems Inc., 3033 Beta Avenue, Burnaby BC
Canada V5G 4M9}
\author{Fabio Altomare}
\affiliation{D-Wave Systems Inc., 3033 Beta Avenue, Burnaby BC
Canada V5G 4M9}
\author{Teresa Medina}
\affiliation{D-Wave Systems Inc., 3033 Beta Avenue, Burnaby BC
Canada V5G 4M9}
\author{Rahul Deshpande}
\affiliation{D-Wave Systems Inc., 3033 Beta Avenue, Burnaby BC
Canada V5G 4M9}
\author{Sara Ejtemaee }
\affiliation{D-Wave Systems Inc., 3033 Beta Avenue, Burnaby BC
Canada V5G 4M9}
\author{Emile Hoskinson}
\affiliation{D-Wave Systems Inc., 3033 Beta Avenue, Burnaby BC
Canada V5G 4M9}
\author{Michael Babcock}
\affiliation{D-Wave Systems Inc., 3033 Beta Avenue, Burnaby BC
Canada V5G 4M9}
\author{Mohammad H.~Amin }
\affiliation{D-Wave Systems Inc., 3033 Beta Avenue, Burnaby BC
Canada V5G 4M9}
\affiliation{Department of Physics, Simon Fraser University, Burnaby, BC Canada V5A 1S6}
\begin{abstract}

We report on measurements of flux and charge noise in an rf-SQUID flux qubit using macroscopic resonant tunneling (MRT). We measure rates of incoherent tunneling from the lowest energy state in the initial well to the ground and first excited states in the target well. The result of the measurement consists of two peaks. The first peak corresponds to tunneling  to the ground state of the target well, and is dominated by flux noise. The second peak is due to tunneling to the excited state and is wider due to an intrawell relaxation process dominated by charge noise. We develop a theoretical model that allows us to extract information about flux and charge noise within one experimental setup. The model agrees very well with experimental data over a wide dynamic range and provides parameters that characterize charge and flux noise. 

\end{abstract}

\maketitle

\section{Introduction}

Improving the performance of superconducting quantum computing technologies relies on reducing the impact of noise sources that lead to decoherence~\cite{de2021materials}. This can be achieved by designing noise-resistant circuits and developing lower-loss materials~\cite{yan2016flux,nguyen2019high,place2021new,siddiqi2021engineering}. The dominant noise sources affecting superconducting qubits are flux and charge noise, which are thought to originate from ensembles of microscopic systems manifesting as materials defects. For qubits implemented with a superconducting quantum interference device (SQUID), a ubiquitous $1/f$ flux noise spectrum has been observed~\cite{bialczak20071,lanting2009geometrical,quintana2017observation,braumuller2020characterizing}. Although a concrete microscopic mechanism for flux noise has yet to be determined, the prevailing models suggest that randomly oriented electronic spins at the metal-oxide interface lead to inductive losses~\cite{koch2007model,anton2013magnetic,lanting2014evidence,lanting2020probing}. Similarly, defects in dielectrics are thought to cause dielectric losses by coupling to and extracting energy from the qubit's electric field~\cite{martinis2005decoherence,muller2019towards}. To design next generation hardware, it is crucial to be able to distinguish and quantify the strength of each noise source in current hardware~\cite{braumuller2020characterizing}.

Macroscopic resonant tunneling (MRT) uses flux-tunable, multi-well qubits to measure the noise affecting the flux qubits~\cite{harris2008probing}. An MRT experiment consists of measuring the incoherent tunneling rate between the flux states of the left and right wells of the qubit potential as a function of flux bias $\efb$ (see Fig. \ref{fig1}). When the energy levels are aligned, the observed MRT peak is shaped by details of the noise spectral density. Previous work on MRT primarily focused on the details of the peak originating from the tunneling between the lowest energy levels in the initial and target wells, which is dominated by flux noise~\cite{harris2008probing,lanting2011probing}. In this article, we report measurements of the lowest energy transition (zeroth peak) and the first excited transition (first peak) corresponding to incoherent tunneling to the first excited state within the target well. Intrawell relaxation from the first excited state to the ground state inside the target well leads to an additional broadening of the first peak. This intrawell relaxation is dominated by charge noise and allows us to characterize the strength of coupling to charge fluctuations. To extract information on the noise affecting the qubits, we develop a theoretical model that combines interwell and intrawell relaxation and takes into account both charge and flux noise that can be fit to experimental data.

\section{\label{section1}Theoretical model}

\begin{figure}
\begin{center}
\includegraphics[width=\columnwidth]{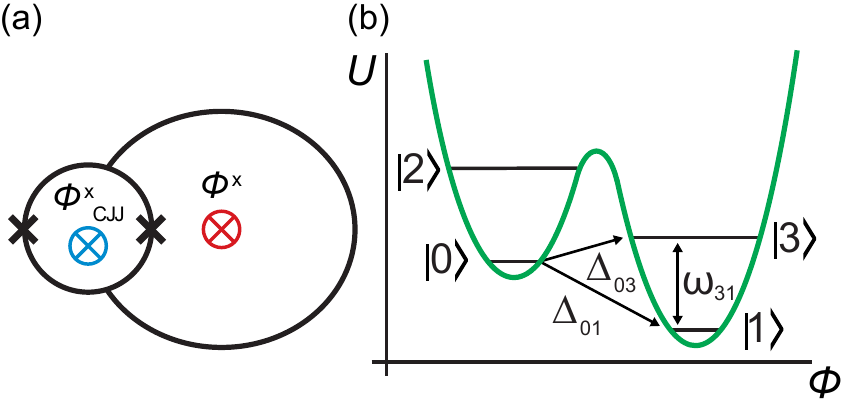}
\caption{\label{fig1} (a) Schematic diagram of an rf-SQUID flux qubit with two external fluxes, $\efb$ and $\ecjj$, threading the two loops. (b) Classical potential energy of the rf-SQUID flux qubit versus the flux $\Phi$ induced in the main loop. Four metastable energy levels are shown, two in each well.}
\end{center}
\end{figure}

We consider a compound Josephson junction (CJJ) rf-SQUID flux qubit, schematically represented in Fig.~\ref{fig1}a \cite{harris2010experimental}. The qubit consists of two loops, main and CJJ, threaded by two external fluxes, $\Phi^x$ and $\Phi^x_{\rm CJJ}$. The potential energy of the qubit has double-well shape as a function of flux $\Phi$ threading the main loop (Fig.~\ref{fig1}b). When the barrier is high, tunneling amplitude between the two wells is small. This allows us to define metastable states $\ket{n}$ with energies $E_n$ in each well, and introduce tunneling amplitudes $\Delta_{mn}$ between states $\ket{m}$ and $\ket{n}$ in opposite wells, as described in Appendix \ref{appendixA}. The Hamiltonian of the system in this basis is written as
\be \label{SHS}
H_{\rm S} = \sum_n E_n \ket{n}\bra{n} -{1\over 2} \sum_{m\neq n} \Delta_{mn} \ket{m}\bra{n}.
\ee 
We numerate states in the left (right) well with even (odd) integers (see Fig.~\ref{fig1}b). For simplicity, we assume in the following that we tunnel from the left initial state $\ket{0}$ into the right well.

The rf-SQUID is dominantly coupled to flux and charge noise via current $I$ through the main loop and voltage $V$ across the junctions, with an interaction Hamiltonian 
\ba \label{SHint}
H_{\rm int} = -\sum_{m,n} (I_{mn} \delta \Phi + V_{mn} \delta q) \ket{m}\bra{n},
\ea
where $I_{mn} = \bra{m}I\ket{n}$ and $V_{mn} = \bra{m}V \ket{n}$. The flux noise, $\delta \Phi$, and charge noise, $\delta q$, are characterized by noise spectral densities $S_\Phi(\omega)$ and $S_q(\omega)$, respectively. Flux noise is taken to be a sum of low-frequency and high-frequency components: $S_\Phi(\omega) = S^\Phi_L(\omega) + S^\Phi_H(\omega)$. The low-frequency part is characterized by its r.m.s. value $W_\Phi$, and the high frequency component is assumed to be ohmic parameterized by a dimensionless parameter $\eta$. Also, charge noise is described by dielectric loss tangent $\tan \delta_C$. Details of the spectral densities and the noise parameters are provided in Appendix \ref{AppendixB}.

At time $t=0$, the system is initialized in the lowest energy state of the left well, $\ket{0}$, with probability $P_0=1$. The rate of transition out of this initial state is given by
\be
\Gamma(\epsilon) = -\left[{dP_0 \over dt} \right]_{t=0} = \sum_n \Gamma_{0n}(\omega_{01} - \omega_{n1}), \label{SGm}
\ee
where $\omega_{n1} = (E_n - E_1)/\hbar$, and $\epsilon = E_0 - E_1 = \hbar \omega_{01}$ is the energy bias from the degeneracy point. The functions $\Gamma_{0n}(\omega)$ describe resonant tunneling from $\ket{0}$ to the state $\ket{n}$ in the target well.

While flux noise directly affects the transition between the wells, charge noise broadens the transition peak indirectly via intrawell relaxation. When the state $\ket{n}$ is an excited state in the target well, after tunneling, the system will quickly relax down to the lowest energy state, $\ket{1}$. The energy uncertainty due to this intrawell relaxation leads to an additional broadening of the transition peak, with a width proportional to the rate of relaxation $\Gamma_{n1}(\omega)$ from $\ket{n}$ to $\ket{1}$. The total transition rate is described by a convolution of three functions, each corresponding to one component of noise (see Appendix \ref{AppendixC}):
\be
\Gamma_{0n}(\omega) = {\Delta_{0n}^2 \over 4 \hbar^2}\, G_{0n}(\omega), \label{SGm0n}
\ee
where
\ba
G_{0n}(\omega) = \!\!\! \int \! {d\omega' \over 2\pi} {d\omega'' \over 2\pi} G_L(\omega {-} \omega')G_H(\omega' {-} \omega'')G_R(\omega''). \quad \label{SG0n}
\ea
Here, we have defined single-peaked functions
 \ba
 G_{L}(\omega) &=& {\sqrt{2\pi}\hbar \over W}\exp \left\{ {-(\hbar\omega-\epsilon_p)^2 \over 2W^2} \right\}, \label{SGL} \\
G_{H}(\omega) &=& {2 \hbar\gamma  \over \hbar^2\omega^2 + \gamma^2}
{\hbar\omega/k_BT   \over 1 - e^{-\hbar\omega/k_BT}}, \label{SGH} \\
G_R(\omega) &=& {2\Gamma_{n1}(\omega+\omega_{n1}) \over \omega^2 + \Gamma_{n1}^2 (\omega+\omega_{n1})}. \label{SGR}
\ea
The Gaussian function $G_{L}(\omega)$ represents 
broadening due to low-frequency flux noise \cite{amin2008macroscopic,harris2008probing}. The width of the peak, $W$, is proportional to the r.m.s. value of low-frequency flux noise. The shift, $\epsilon_p$, is related to $W$ by the fluctuation-dissipation theorem: 
\be
W^2 = 2k_BT \epsilon_p.
\label{FDT}
\ee
High frequency flux noise is included via the Lorentzian-like function $G_H(\omega)$, with broadening determined by $\gamma$ \cite{amin2008macroscopic,smirnov2018theory}. Finally, intrawell relaxation is captured by $G_R(\omega)$. The function $\Gamma_{n1}(\omega)$ represents intrawell relaxation from state $\ket{n}$ to $\ket{1}$ within the target well. Naturally for the lowest MRT peak with $n=1$, there is no intrawell relaxation. Therefore, $\Gamma_{11} = 0$ and $G_R(\omega) = 2\pi\delta(\omega)$, turning Eq.~\eqref{SGm0n} into a single convolution integral: 
\ba
G_{01}(\omega) = \!\!\! \int {d\omega' \over 2\pi}G_{L}(\omega {-} \omega')G_{H}(\omega'). \quad \label{fit2}
\ea
For transition to the first excited state in the target well with $n=3$, we approximately write (see Appendix \ref{AppendixC})
\ba
\Gamma_{31}(\omega) &=&  {\zeta \over \hbar} { \tanh (\hbar\omega/k_BT) \over 1 - e^{-\hbar\omega/k_BT}}, \label{SGmn1} 
\ea 
where $\zeta$ is a charge noise broadening coefficient in units of energy. For multi-level MRT peaks with $n>3$, there are more than one intrawell relaxation channels, and $\Gamma_{n1}$ is the sum of all of them. In this paper, however, we only focus on the first two MRT peaks. This model is an extension to previous models that relied on a convolution of two noise sources \cite{amin2008macroscopic,smirnov2018theory}. A more formal derivation of the model introduced here can be found in [\!\!\citenum{theorypaper}]. Note that all  $G_{\kappa}(\omega)$ functions, with $\kappa = L,H,R, 0n$, are approximately normalized (with slight deviations due to non-ideal Lorentzian form) with normalization condition,
\be
\int {d\omega \over 2\pi} G_\kappa(\omega) = 1. \label{Snorm}
\ee
Any deviation from a perfect normalization will be absorbed into $\Delta_{0n}$, when taken as a free parameter.

The parameters $\Delta_{0n}$, $W$, $\gamma$, and $\zeta$ are all in units of energy and can be calculated using the underlying rf-SQUID Hamiltonian and noise spectral densities as described in the appendices. The broadening parameters $W$, $\gamma$, $\zeta$ would then depend on the target state $\ket{n}$ and the energy bias $\epsilon$. Ignoring these small dependencies and under some additional assumptions listed in Appendix \ref{AppendixD}, we can treat them as fitting parameters. This allows us to fit the model to the experimental data and extract information about noise with no need for diagonalization of the rf-SQUID Hamiltonian. The energy bias $\epsilon$ at which MRT peaks are measured is obtained by applying the external flux $\Phi^x$. This bias can be approximated by $\epsilon = 2 I_P \Phi^x$, where $I_P = (I_{11} - I_{00})/2$ is the persistent current, and $\Phi^x$ is measured from the degeneracy point $\Phi_0/2$. 
One can therefore present the transition rate as a function of the applied flux, $\Gamma(\Phi^x)$, and express the noise parameters in flux units
\be
W_\Phi = {W \over 2I_P}, \quad 
\epsilon_\Phi = {\epsilon \over 2I_P}, \quad
\gamma_\Phi = {\gamma \over 2I_P}, \quad
\zeta_{\Phi} = {\zeta \over 2 I_P}.
\ee
While $W_\Phi$ directly measures the r.m.s. value of low-frequency flux noise, the dimensionless ohmic coefficient of high-frequency flux noise, and the loss tangent of charge noise are given by (see Appendix \ref{AppendixD})
\be
\eta = {4I_P \gamma_\Phi \over k_BT}, \qquad 
\tan \delta_C = {\zeta_\Phi \over \Phi^x_{31}}, \label{etatanC}
\ee
where $\Phi^x_{31} = \hbar \omega_{31}/2I_P$ is the distance between the two peaks in flux units. It is also common to express ohmic flux noise in terms of shunt resistance $R_S$ or inductive loss tangent $\tan \delta_L$ \cite{nguyen2019high}:
\be
 R_S = {8I_P^2L^2 \over \hbar \eta} = {2I_P L^2 k_BT \over \hbar \gamma_\Phi}, \qquad
\tan \delta_L =  {\omega L \over R_S},
\label{shuntresistance}
\ee
where $L$ is the inductance of the main loop.

\begin{figure}
\begin{center}
\includegraphics[width=\columnwidth]{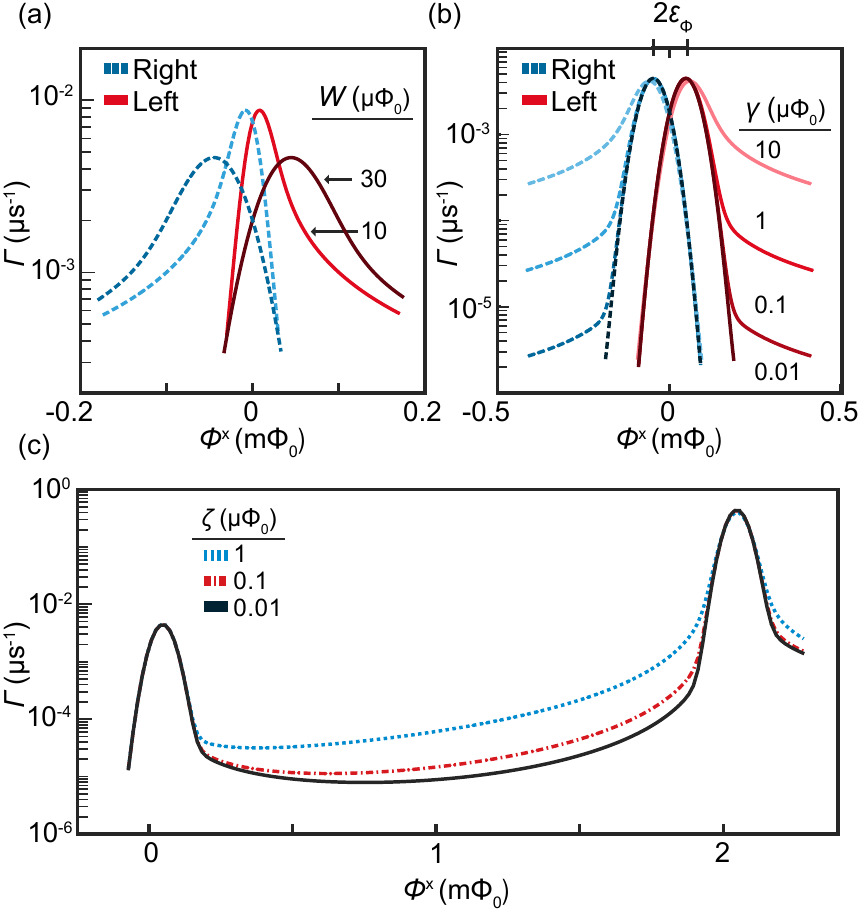}
\caption{\label{fig2} Macroscopic resonant tunneling rate as a function of external flux $\Phi^x$ (measured from the degeneracy point $\Phi_0/2$) for different values of (a) low-frequency flux noise broadening $W_\Phi$, (b) high-frequency flux noise broadening $\gamma_\Phi$, and (c) charge noise broadening $\zeta_\Phi$.  Red and blue line-shapes in (a) and (b) correspond to zeroth peak with left and right well initialization, respectively. The line-shapes in (c) represent zeroth and first peaks corresponding to the left initialization at a constant $\Phi^x_{\rm CJJ}$. We use $\Delta_{01}=2$~MHz and $ T=5$~mK in all simulations. We also use $\gamma_\Phi = 10~\mu\Phi_0$ for (a), $W_\Phi = 35~\mu\Phi_0$ for (b), and  $\Delta_{03}=20$~MHz, $W_\Phi = 35~\mu\Phi_0$, $\gamma_\Phi = 3~\mu\Phi_0$ for (c).}
\end{center}
\end{figure}

To illustrate the dependence of the MRT peaks on the broadening parameters, we plot the transition rate $\Gamma(\Phi^x)$ for different $W_\Phi$,  $\gamma_\Phi$, and $\zeta_\Phi$ in Fig.~\ref{fig2}. Figure \ref{fig2}a,b highlights the zeroth peaks for both left and right well initial state preparations (colored as red and blue respectively). In Fig.~\ref{fig2}a, the peaks are plotted for two  values of the width: $W_\Phi = 10$ and $30\,\mu\Phi_0$. The shift of the peak position from zero bias is $\epsilon_\Phi$, which is also changed according to \eqref{FDT} assuming constant temperature \cite{harris2008probing}. Figure \ref{fig2}b shows MRT peaks for different strengths of high frequency noise $\gamma_\Phi$ while keeping all other parameters constant. For small values $\gamma_\Phi$, low-frequency flux noise dominates, and a Gaussian broadened line shape of width $W_\Phi$ is recovered~\cite{amin2008macroscopic,harris2008probing}. When $\gamma_\Phi$ increases, an additional broadening develops with a characteristic asymmetric tail extending to larger flux biases~\cite{lanting2011probing}. 
While intrawell relaxation does not contribute to the broadening of $\Gamma_{01}$ in Eq.~\eqref{SGm}, it does for higher energy transitions such as $\Gamma_{03}$ as shown in Fig.~\ref{fig2}c. For increasing values of the broadening parameter $\zeta_\Phi$, i.e., higher intrawell relaxation, the width of the first peak and its contribution to the valley between the two peaks are increased.

\section{\label{section2}Experimental Results}

The MRT measurement protocol involves preparing the qubit in a known initial state of the rf-SQUID double-well potential (ground state of the left or right well) and measuring the tunneling rate into the adjacent well as a function of flux bias as described in [\!\!\citenum{harris2008probing}]. MRT measurements were performed on the quantum processor of a D-Wave 2000Q\textsuperscript{\texttrademark} lower noise system. The qubit has external lines that apply fluxes $\efb$ and $\ecjj$ to the main and CJJ loops, respectively. These lines enable time-dependent control over the qubit potential energy, with $\efb (t)$ setting the flux-bias tilt between the left and right well, and $\ecjj(t)$ tuning the tunneling energy $\Delta$. Each qubit is controlled by external lines that have a 3 and 30 MHz bandwidth, respectively, that enable \textit{in-situ} MRT measurements on individual qubits throughout the quantum processor.

We measured 27 parametrically identical qubits across the fabric of the processor. The measurements were performed in a dilution refrigerator with a base temperature of $\!\sim$10~mK on a processor fully calibrated according to the procedure described in [\!\!\citenum{harris2010experimental}].  The qubits had a critical current of $I_c  = 2.30 \pm 0.08 ~\mu{\rm A}$, an inductance of $L = 250 \pm 7 ~ {\rm pH}$ and a capacitance of $C = 110 \pm  4 ~{\rm fF}$. A constant bias of $\ecjj = - 0.74~ \Phi_0$ was applied to facilitate measurements of the tunneling rate that varied over four orders of magnitude as a function of $\efb$. At this value of $\ecjj$, the qubits had a persistent current of $I_p =  1.37 \pm 0.01~\mu{\rm A}$.

\begin{figure}
\begin{center}
\includegraphics[width=\columnwidth]{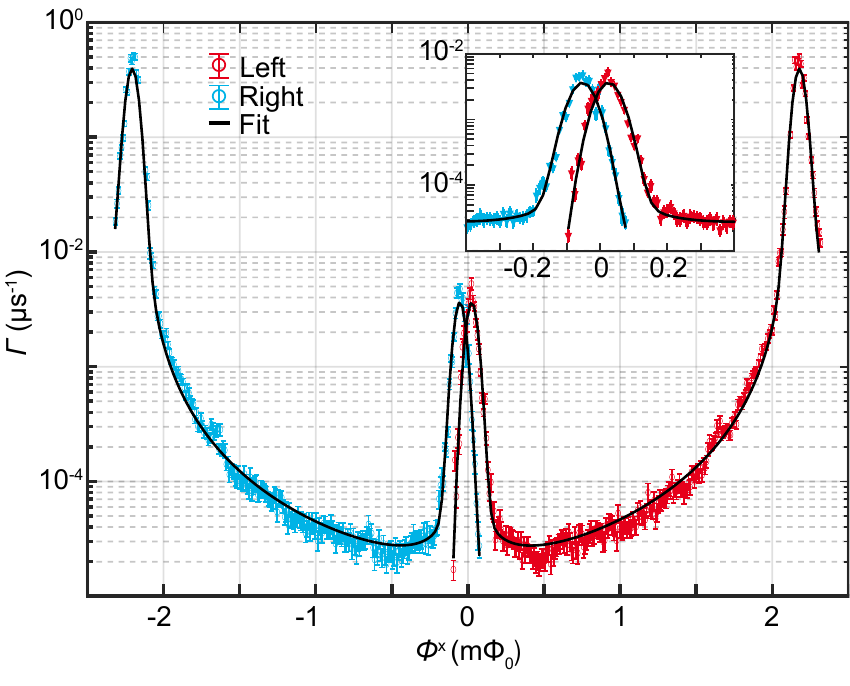}
\caption{\label{fig3} Measurement of macroscopic resonant tunneling rate $\Gamma$ as a function of flux bias $\efb$ for a single qubit at a fixed $\ecjj$ controlling the barrier height. The red (blue) colour shows the initial state prepared in the left (right) well. The solid line is a fit to the model described in Eqs.~\eqref{fit1}. The inset shows the zeroth MRT peaks, which highlights the asymmetric tail resulting from high-frequency flux noise. The first MRT peak has additional broadening due to intrawell relaxation, which allows the extraction of the strength of charge noise.}
\end{center}
\end{figure}

Figure \ref{fig3} shows a typical dataset of a single qubit's tunneling rate as a function of flux bias with an initial state preparation in the left (right) well shown in red (blue). At the degeneracy point, $\efb = 0$, the ground states of the two wells are aligned. The data shows a resonant peak near this point that corresponds to tunneling between these two states with an offset depending on the state initialization (see Eq. \ref{FDT} and [\!\!\citenum{harris2008probing}]). This zeroth MRT peak has a width dominated by low-frequency flux noise. Away from this peak, the tunneling rate exhibits an asymmetric tail representative of high-frequency flux noise, in qualitative agreement with Fig.~\ref{fig2}b. Further increasing $|\efb|$ causes a gradual increase in the tunneling rate until reaching the first peak at $|\efb| \sim 2 ~\rm{m}\Phi_0 $. At this point, the initial state is aligned with the first excited state in the target well. An additional broadening is observed on the first peak due to intrawell relaxation in the target well. The line-shape near this peak and in the valley between the two peaks provides information about the strength of the charge noise.

We fit the model described in Eqs.~\eqref{SGm}-\eqref{SGmn1} to the measured tunneling rates with,
\be
\Gamma (\efb) = {1 \over 4} \big(\Delta_{01}^2 G_{01}(\efb) + \Delta_{03}^2 G_{03}(\efb)\big), \label{fit1}
\ee
where $G_{01}(\efb)$ and $G_{03}(\efb)$ are described by Eq. \ref{fit2} and \ref{SG0n}.

A typical best fit to the dataset is shown by the solid black line in Fig.~\ref{fig3} with the tunneling amplitudes $\Delta_{01} / h = 2.72 \pm 0.01~ \rm{MHz}$ and $\Delta_{03} / h = 29.8 \pm 0.2~\rm{MHz} $, and the noise broadening parameters $W_\Phi = 37.2 \pm 0.1~\mu \Phi_0$, $\gamma_\Phi = 0.54 \pm 0.05~\mu \Phi_0$, and $\zeta_\Phi = 4.53 \pm  0.09~\mu \Phi_0 $. Flux offset $\Phi^x_{31} = 2153.6  \pm 0.5~\mu \Phi_0$, corresponding to $\omega_{31}/2\pi = 2I_P\Phi^x_{31}/h =  9.17$ GHz, is used to fit the relative position of the zeroth and first peaks. The fitting temperature of $T = 7.3$~mK matches the thermometry mounted on the mixing chamber plate. Using \eqref{etatanC} and \eqref{shuntresistance} we estimate the noise parameters: $\eta=5.9\pm0.5\times 10^{\rm{-2}}$, $R_{\rm S} = 147 \pm 13~\rm{k}\Omega$, $\tan \delta_C = 2.07 \pm 0.04 \times 10^{\rm{-3}}$, and $ \tan \delta_L(1\,\text{GHz}) = 10.6 \pm 0.9 \times 10^{\rm{-6}}$.

\begin{figure}
\begin{center}
\includegraphics[width=\columnwidth]{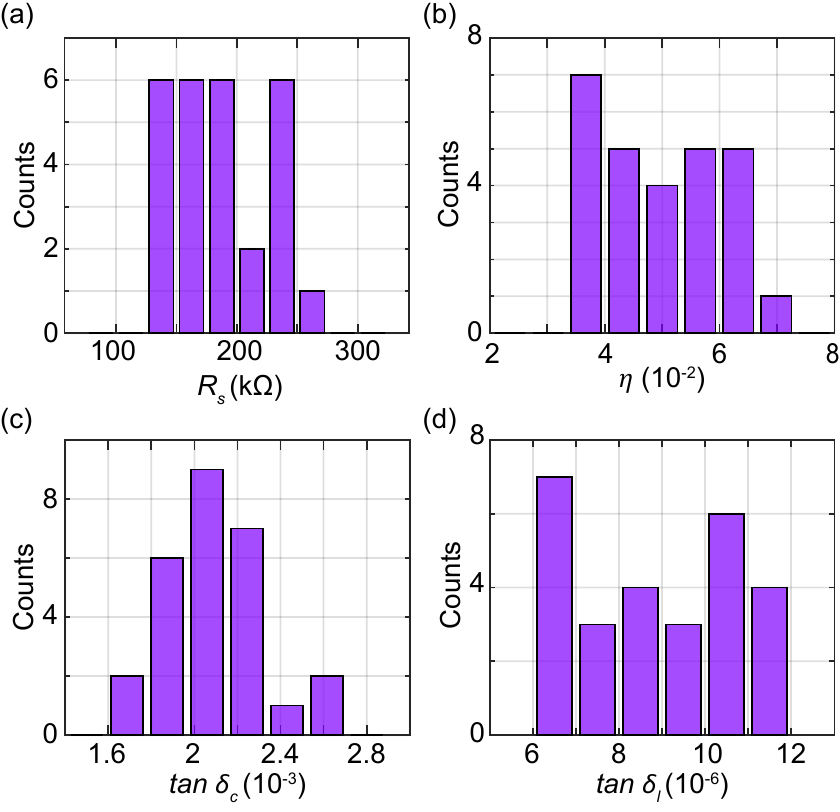}
\caption{\label{fig4} Variation in noise parameters for the 27 qubits calculated using \eqref{etatanC} and \eqref{shuntresistance} with $\tan \delta_L(1\,\text{GHz})$ in (d). The parameters are calculated from fits to data sets that are similar to Fig. \ref{fig3}.}
\end{center}
\end{figure}

Using the same measurement procedure we fit all 27 qubits to the hybrid noise model and find consistent results. The data and fit for each of these qubits is similar to Fig. \ref{fig3}. A summary of these results is presented in Fig.~\ref{fig4}. We find mean noise parameters of $\eta=5\pm1 \times 10^{\rm{-2}}$, $R_{\rm S} = 180 \pm 40~\rm{k}\Omega$, $\tan \delta_C = 2.1 \pm 0.2 \times 10^{\rm{-3}}$, and $ \tan \delta_L(1\,\text{GHz}) = 8.9 \pm 1.8 \times 10^{\rm{-6}}$. The extracted $\tan \delta_C $ is consistent with the expected value for amorphous SiO$_{\rm{x}}$ and the low and high frequency flux noise is similar to previous experiments \cite{lanting2011probing}.

The approximate model used to fit the data of Fig. \ref{fig3} is obtained under the assumptions listed in appendix \ref{AppendixD}. A more accurate model uses the CJJ rf-SQUID Hamiltonian \eqref{HS} to calculate the tunneling amplitudes and noise broadening parameters, as described in appendices \ref{appendixA} to \ref{AppendixC} and also in [\!\!\citenum{theorypaper}].
In Fig.~\ref{fig5}, we compare the simplified model to the full model using the same noise parameters found for Fig.~\ref{fig3} and the qubit parameters reported above. We find an overall good agreement between the two models, with the simplified model resulting in a $\sim$5\% better $\chi^2$ value due to uncertainty in qubit parameters. The agreement between the models gives us confidence in the noise parameters extracted from the approximate model, which requires significantly less computational resources.

\section{Conclusion}

We introduce a hybrid noise model for macroscopic resonant tunneling in rf-SQUID flux qubits. The model includes contributions of low and high frequency flux noise as well as charge noise. We fit the experimentally measured MRT rates to the model and find good agreement over a dynamic range of four orders of magnitude. Each noise component generates a characteristic line-shape broadening that is captured by the fit. This allows the noise sources to be uniquely identified and quantified. The ability to extract information about different sources of noise in a single experimental setting and {\it in-situ} on the quantum processor is an important step towards understanding the origin of the measured noise and providing an indication on how to reduce it. This will ultimately be crucial for the development of quantum computers.

\begin{acknowledgments} 

We acknowledge fruitful discussions with Richard Harris. We also thank Joel Pasvolsky for a careful reading of the paper.

\end{acknowledgments}

\appendix

\section*{Appendix}
In the following appendices we provide details of the theoretical model used in the main text. While in the main text we have used the full expressions, for simplicity we will use $\hbar = k_B = 1$ in the appendices.

\section{rf-SQUID Hamiltonian}
\label{appendixA}

A simplified version of a compound Josephson junction (CJJ) flux qubit \cite{harris2010experimental} is sketched in Fig.~\ref{fig1}(a). It has two superconducting loops, main and CJJ loops, with flux degrees of freedom $\Phi$ and $\Phi_{\rm CJJ}$, subject to external flux biases $\efb$ and $\ecjj$, respectively. The Hamiltonian of such
an rf-SQUID is written as
 \ba
 H_S
 = {q^2 \over 2C} + {q_{\rm CJJ}^2 \over 2C_{\rm CJJ}} + U(\Phi,\Phi_{\rm CJJ}) \label{HS}
 \ea
where $C$ and $C_{\rm CJJ}$ are parallel and series combinations of the
junction capacitances, $q$ and $q_{\rm CJJ}$ are the sum and difference of
the charges stored in the capacitors respectively, and
 \ba
 U(\Phi,\Phi_{\rm CJJ}) =  \frac{(\Phi{-}\efb {+} \Phi_0/2)^2}{2L}
 + \frac {(\Phi_{\rm CJJ}{-}\ecjj)^2}{2L_{\rm CJJ}} \nn
 - E_J \cos \left(\frac{\pi \Phi_{\rm CJJ}}{\Phi_0}\right)\cos\left(\frac{2\pi \Phi}{\Phi_0}\right)
 \qquad \label{U}
 \ea
is a 2-dimensional potential with $L$ and $L_{\rm CJJ}$ being the inductances of the
two loops, and $\Phi_0 {=}\, h/2e = \pi \hbar/e$ is the flux quantum. We have assumed
symmetric Josephson junctions forming the CJJ loop, with  a total critical current $I_C = 2\pi E_J/\Phi_0$ through both junctions.  Flux and charge degrees of freedom satisfy commutation relations: $[\Phi,q] = [\Phi_{\rm CJJ}, q_{\rm CJJ}] = i\hbar$. 
 
The CJJ loop typically has a small inductance,  $L_{\rm CJJ} \ll L$, making dynamics of $\Phi_{\rm CJJ}$ much faster than $\Phi$. Therefore, the qubit's quantum properties is dominantly determined by tunneling in the $\Phi$ direction. The environment also mostly affects the qubit via $\Phi$ and $q$ degrees of freedom. We therefore write the interaction Hamiltonian as
 \be \label{hint1}
 H_{\rm int} = -{\Phi - \efb +\Phi_0/2 \over L} \,\delta\Phi - {q \over C} \,\delta q \equiv - I \delta \Phi - V \delta q,
 \ee
where $\delta \Phi$ and $\delta q$ are flux and charge noise operators, respectively, and
\be \label{cv1}
I = {\Phi - \efb +\Phi_0/2 \over L}, \qquad V = {q \over C} = -{i\hbar \over C} {\partial \over \partial \Phi}
\ee
are loop current and junction voltage operators. The flux and charge noises are described by noise spectral densities
\ba \label{spectrum}
S_\Phi(\omega) \!\!&=&\!\!  \int dt \, e^{i\omega t}\langle \delta \Phi(t)\delta \Phi(0)\rangle , \\
S_q(\omega) \!\!&=&\!\!  \int dt \, e^{i\omega t}\langle \delta q(t)\delta q(0)\rangle. 
\ea
where $\langle \dots \rangle$ represents averaging over environmental degrees of freedom.

Experiments are performed when $U(\Phi,\Phi_{\rm CJJ})$ forms a double-well potential along the $\Phi$ direction, with a large barrier between the wells. The lowest energy states in each well are then metastable with small amplitudes of tunneling to states in the opposite well.  At the degeneracy point, $\Phi^x = 0$, the minima in the two wells align. We follow the procedure described in [\!\!\citenum{amin2013adiabatic}] to determine the metastable states $\ket{n}$. We divide the Hilbert space into two subspaces with $\Phi {-} \Phi^x {+} \Phi_0/2 < 0 $ and $\Phi {-} \Phi^x {+} \Phi_0/2 > 0$ corresponding to the two wells. We then partially diagonalize the Hamiltonian in each subspace to determine $\ket{n}$. The system Hamiltonian \eqref{HS} can now be written in this new basis as
\be \label{h32}
H_{\rm S} = \sum_n E_n \ket{n}\bra{n} - {1 \over 2} \sum_{m\neq n} \Delta_{mn} \ket{m}\bra{n},
\ee 
where
\ba \label{en1}
E_n = \langle n | H_S |n\rangle, \qquad \Delta_{mn} = -2\langle m | H_S |n \rangle .
\ea
The interaction Hamiltonian in this representation becomes
\ba \label{hint2}
H_{\rm int} = -\sum_{m,n} (I_{mn} \delta \Phi + V_{mn} \delta q) \ket{m}\bra{n},
\ea
where 
\ba  \label{IVnm}
I_{mn} = \bra{m}I\ket{n}, \qquad V_{mn} = \bra{m}V \ket{n}.
\ea
We numerate states in the left (right) well with even (odd) integers (see Fig.~\ref{fig1}b). Due to the construction of the basis, we have $\Delta_{mn} = 0$ within each well (between two even or two odd states) and $I_{mn} = 0$ for every pair of states in opposite wells (for odd $m{+}n$). Also, since $\ket{n}$ is delocalized in charge, we expect $V_{nn} = 0$ for all $n$. We define persistent current $I_P$ as the expectation value of current in the lowest state of each well when the rf-SQUID is at the degeneracy:
\ba \label{ip1}
I_P = \frac{I_{11} - I_{00}}{2}. 
\ea
In principle, persistent current is bias-dependent, but the dependence is expected to be weak. 

\section{Noise spectral density}
\label{AppendixB}

Both flux noise and charge noise affect the shape of resonant tunneling peaks. In principle, frequency dependence of flux noise can be different at low and high frequencies. We therefore write flux noise as a sum of two components,
\be
S_\Phi(\omega) = S^\Phi_L(\omega) + S^\Phi_H(\omega),
\ee
with different frequency dependencies. The low-frequency component typically has $1/f$ type of spectrum
\be
S^\Phi_L(f) = {A_\Phi^2 \over |f/f_0|^{\alpha}}
\ee
where $f_0 = 1\,$Hz, $\alpha \lesssim 1$, and $A_\Phi$ is typically of order of a few $\mu \Phi_0/\sqrt{Hz}$. In practice, low-frequency noise affects the MRT line-shape through its rms value
\be
W_\Phi = \sqrt{ \int {d\omega \over 2\pi} S^\Phi_L (\omega)},
\ee
which can also be expressed in units of energy
\be
W = 2 I_P W_\phi.
\ee
The relation between $W_\Phi$ and $A_\Phi$ can be non-trivial. Modeling this relation requires knowledge of accurate noise frequency dependence at intermediate frequencies and proper introduction of integration bounds. We therefore take $W_\Phi$ directly as an independent fitting parameter in our model.

At high frequencies, flux noise is typically ohmic \cite{lanting2011probing} with spectral density
\be
S^\Phi_{H} (\omega) = {1 \over 4I_P^2} {\eta \omega  e^{-|\omega|/\omega_c} \over 1 - e^{-\omega/T}},
\label{Ohmic}
\ee
where $\eta$ is a dimensionless coupling coefficient and $\omega_c$ is a cutoff frequency. We assume $\omega_c$ is larger than all relevant frequencies and ignore it. We expect $S_\Phi(\omega)$ to be determined only by flux noise and not by the qubit's operation point. This means $S_\Phi(\omega)$, and therefore $W_\Phi$, are independent of $I_P$, hence
\be
W \propto I_P, \qquad \eta \propto I_P^2.
\ee
To have a quantity that is independent of $I_P$, we introduce inductive loss tangent, $\tan \delta_L$, via
\be \label{SPhisgn}
S_\Phi(\omega) = 2L \, {{\rm sgn}(\omega) \tan \delta_L(\omega) \over 1 - e^{-\omega/T}}.
\ee
The ${\rm sgn}(\omega)$ is added to make $\tan \delta_L$ positive. To agree with $S_\Phi(\omega)$ in both low and high frequency regimes, we should have
\ba
\tan \delta_L(\omega) \!=\! \left\{
\begin{array}{cc}
   (2\pi f_0)^\alpha (A_\Phi^2 / 2LT) |\omega|^{1-\alpha},  &  |\omega| \ll T \\
   & \\
   (\eta/8L I_P^2) |\omega|,   & |\omega| \gtrsim T
\end{array} 
\right. \! . \ \ \
\ea
Notice that $\tan \delta_L$ is independent of $I_P$. It is also insensitive to the rf-SQUID geometry if $S_\Phi \propto L$, which is the case if the length of the qubit wire is changed without changing its width \cite{lanting2009geometrical}.

Similar to flux noise, charge noise also  has $1/f$ spectral density at low-frequencies
\be \label{Sqf}
S_q(f) = {A_q^2 \over |f/f_0|^{\alpha_q}}, \qquad f_0 = 1\, \text{Hz}
\ee
with $\alpha_q \approx 1$ and $A_q \sim 10^{-2}$ - $10^{-4} \, e/\sqrt{\text{Hz}}$. The $1/f$ spectral density typically crosses over to a different frequency dependence at higher frequencies, which is likely to be ohmic \cite{Astafiev,Oliver}. It is common to express charge noise in terms of a capacitive loss tangent, $\tan \delta_C$, that characterises the quality of the dielectrics and two-level fluctuators in the environment, independent of the qubit. We therefore define
\be \label{Sqsgn}
S_q(\omega) = 2C \, {{\rm sgn}(\omega) \tan \delta_C \over 1 - e^{-\omega/T}}.
\ee
The sign function ${\rm sgn}(\omega)$ is needed to make the numerator antisymmetric while keeping $\tan \delta_C$ positive. At low frequencies, $S_q(\omega) \to 2CT \tan \delta_C/|\omega|$, which reduces to \eqref{Sqf} with $\alpha_q=1$ if
\be
\tan \delta_C = 2\pi {E_C \over T} \left({A_q \over e/\sqrt{\text{Hz}}} \right)^2
\ee
where $E_C = e^2/2C$ is the charging energy. At high frequencies, \eqref{Sqsgn} leads to $S_q(\omega) \propto \tan \delta_C$, which requires $\tan \delta_C \propto \omega$ for ohmic spectral density. We therefore expect $\tan \delta_C$ to be constant at low frequencies with a crossover to a different frequency dependence at high frequencies.

\section{Macroscopic resonant tunneling}
\label{AppendixC}

Our goal is to calculate the rate of incoherent tunneling between the two wells. Suppose at time $t=0$ the rf-SQUID is initialized in state $\ket{0}$ with probability $P_0=1$. The probability $P_0$ will decrease with time as the system tunnels to states in the opposite well. We define the MRT transition rate as:
\be
\Gamma = -\left[{dP_0 \over dt} \right]_{t=0}.
\ee
In general transition out of state $\ket{0}$ happens via tunneling to more than one state in the target well. We therefore write
\be
\Gamma(\epsilon) = \sum_n \Gamma_{0n}(\epsilon - \omega_{n1}), \label{S1}
\ee
where $\Gamma_{0n}$ is the transition rate from the initial state $\ket{0}$ to state $\ket{n}$ in the target well,  $\epsilon = E_0 - E_1$ is the energy bias from the degeneracy point, and $\omega_{nm} = E_n - E_m$. Each $\Gamma_{0n}$ can be calculated independently. In the next two subsections, we describe the zeroth and first peak $\Gamma_{01}$ and $\Gamma_{03}$.

\subsection{Tunneling between the lowest energy states}

To calculate $\Gamma_{01}$, we need to consider only two states $\ket{0}$ and $\ket{1}$, corresponding to the ground states in the left and right wells, respectively. We can therefore represent the system Hamiltonian in terms of Pauli matrices
\ba \label{pauli1}
\sigma_x &=& \ket{0}\bra{1} + \ket{1}\bra{0}, \nn
\sigma_z &=& \ket{1}\bra{1} - \ket{0}\bra{0}.
\ea
The effective Hamiltonian in this subspace becomes
\ba \label{h0}
H_S = -\frac{\epsilon}{2} \,\sigma_z - \frac{\Delta_{01}}{2}\,\sigma_x.
\ea
It can be shown that
\ba \label{ep7}
\epsilon \approx 2 I_P \, \efb, 
\ea
with the external flux $\efb$ measured relative to the degeneracy point. Substituting the current operator $I = I_P \sigma_z$ into Eqs.~\eqref{hint2}, the interaction Hamiltonian for flux noise becomes
\ba \label{hint3}
H_{\rm int}^\Phi = - I_P \,\delta \Phi \, \sigma_z = - \frac{1}{2}\, Q \,\sigma_z, 
\ea
with the noise operator 
\ba \label{q1-1}
Q = 2 \,I_P\, \delta \Phi. 
\ea
The effective Hamiltonian of the two-state system describing the rf-SQUID coupled to flux noise environment is therefore given by
\ba \label{h3}
H = - \frac{\epsilon}{2} \,\sigma_z - \frac{\Delta_{01}}{2}\,\sigma_x  - \frac{1}{2}\, Q \,\sigma_z.
\ea
We introduce the spectral density corresponding to operator $Q$ as
\ba \label{spec1}
S_Q(\omega) =  \int dt \, e^{i\omega t}\langle Q(t)Q(0)\rangle = 4I_p^2 S_\Phi(\omega).
\ea
Like $S_\Phi(\omega)$, we can decompose this spectral density into low and high frequency components: $S_Q(\omega) = S^Q_L(\omega) + S^Q_H(\omega)$.  It was shown in [\!\!\citenum{smirnov2018theory}] that the MRT transition rate can be expressed by a convolution integral
\ba
\Gamma_{01}(\epsilon) \approx {\Delta_{01}^2 \over 4} \int {d\omega \over 2\pi} G_{L}(\epsilon {-} \omega)G_{H}(\omega).
\label{Conv}
\ea
The effect of low-frequency flux noise is captured by the Gaussian envelope
\be
G_{L}(\omega) = {\sqrt{2\pi} \over W}\exp \left\{ {-(\omega-\epsilon_p)^2 \over 2W^2} \right\},  \label{GL}
\ee
where
\ba
W^2 \!\!&=&\!\! \int {d\omega \over 2\pi} S^Q_L(\omega),  \label{W-2}\\
\epsilon_{p} \!\!&=&\!\! {\cal P} \int {d\omega \over 2\pi} {S^Q_L(\omega) \over \omega}. \label{ep-2}
\ea
Here, ${\cal P}$ represents the principal value integral. Fluctuation-dissipation theorem requires\cite{amin2008macroscopic} 
\be
W^2 = 2T \epsilon_p.
\ee
High frequency noise affects the peak through a Lorentzian-like envelope function
\be
G_{H}(\omega) \approx {S^Q_H(\omega) \over \omega^2 + [{1\over 2}S^Q_H(\omega)]^2},
\label{GH0}
\ee
with
\ba 
S^Q_H(\omega) = {\eta \omega \over 1 - e^{-\omega/T}}.
\ea
At small frequencies, $\omega \ll T$, we have $S^Q_H(\omega) = \eta T$, therefore, the denominator of  \eqref{GH0} can be written as $\omega^2 + \gamma^2$, where $\gamma \equiv \eta T/2$. At large frequencies, $\omega \gg T$, we have $S^Q_H(\omega) = \eta \omega$, and the denominator of \eqref{GH0} becomes $(1+\eta^2/4)\omega^2 \approx \omega^2$ for $\eta \ll 1$. Therefore, we can express the high frequency envelope function in terms of the broadening parameter $\gamma$ as
\be
G_{H}(\omega) \approx {2 \gamma \omega/T \over (\omega^2 + \gamma^2) (1 - e^{-\omega/T})}.
\label{GH}
\ee

The two parameters $W$ and $\gamma$ measure the width of the envelope functions $G_L$ and $G_H$ in energy units, respectively. Each envelope function approximately becomes a delta function when the width goes to zero. In the absence of high frequency noise, $\gamma \to 0$, we have
\ba
\Gamma_{01}(\epsilon) \approx {\Delta_{01}^2 \over 4} G_{L}(\epsilon),
\ea
in agreement with [\!\!\citenum{amin2008macroscopic} and \citenum{harris2008probing}]. Similarly, when low-frequency noise is absent ($W,\epsilon_p \to 0$), the Gaussian function \eqref{GL} becomes $\delta$-function, and we obtain
\be
\Gamma_{01}(\epsilon) \approx {\Delta_{01}^2 \over 4} {S^Q_H(\epsilon) \over \epsilon^2 + \gamma^2}.
\ee
For $\epsilon > \gamma$  we get
\be
\Gamma_{01}(\epsilon) \approx {\Delta_{01}^2 \over 4\epsilon^2} S^Q_H(\epsilon)
\ee
in agreement with the Bloch-Redfield theory. For small biases, we obtain the Lorentzian relaxation rate expected for white noise \cite{amin2008macroscopic}
\be
\Gamma_{01}(\epsilon) \approx {\Delta_{01}^2 \over 2}{\gamma  \over \epsilon^2 + \gamma^2}.
\ee
Therefore, the convolution form \eqref{Conv}, with envelope functions \eqref{GL} and \eqref{GH}, gives correct results in all limiting regimes. It can also be numerically shown that it agrees well with the exact results in other regimes as long as $\eta \ll 1$. One advantage of the convolution form is that it separates contributions of low and high frequency noise into two separate envelope functions. It is therefore possible to study the effect of each noise separately and calculate the corresponding envelope function. We will use this property to determine the effect of intrawell relaxation on multi-level MRT peaks in the next subsection.

\subsection{Tunneling to a higher energy state}
\label{HigherOrder}

We now consider multi-level MRT peaks when tunneling happens to a higher energy state in the target well. For simplicity, we consider transition to the second level in the target well, i.e., $\Gamma_{03}$. As before, we assume that the system is initialized in state $\ket{0}$. Incoherent tunneling from $\ket{0}$ to $\ket{3}$ is affected by the flux noise the same way as discussed in the previous subsection. The broadening due to low-frequency and high-frequency noise is captured by $G_L(\omega)$ and $G_H(\omega)$, respectively, with minor changes to the parameters that we shall mention below. However, since $\ket{3}$ is an excited state within the target well, the system will quickly relax to state $\ket{1}$, in a time scale much shorter than the incoherent tunneling rate. The uncertainty in energy $E_3$ due to the intrawell relaxation creates an additional broadening. Such a broadening was introduced in [~\citenum{amin2008macroscopic}], but the resulting transition peak was symmetric around its center, violating the detailed balance needed to reach Boltzmann distribution in thermal equilibrium.
Here, we provide a simple calculation of the broadening effect in a way that satisfies detailed balance. A more complete derivation is provided in [\!\!\citenum{theorypaper}].

As we mentioned before, the broadening effect of every component of noise can be calculated independently and combined together through a convolution integral. The combined transition rate is
\be
\Gamma_{03}(\epsilon) = {\Delta_{03}^2 \over 4}\, G_{03}(\epsilon - \omega_{31}),
\ee
with
\be
G_{03}(\omega) = \!\!\! \int {d\omega' \over 2\pi} {d\omega'' \over 2\pi} G_L(\omega {-} \omega')G_{H}(\omega' {-} \omega'')G_{R}(\omega''), \quad \label{GammaTotal}
\ee
where $G_L(\omega)$ and $G_H(\omega)$ are given by \eqref{GL} and \eqref{GH}, respectively, and $G_R(\omega)$ is a peaked function capturing the broadening due to the intrawell relaxation. As before, $\epsilon = E_0 - E_1$ is the energy bias measured from the rf-SQUID degeneracy point. We therefore have
\be
\omega_{03} = \epsilon -\omega_{31} \approx 2 I_P^{03} (\efb - \Phi^x_{31}). 
\ee 
Here, we define $\Phi^x_{31}$ as the value of the external flux $\efb$ when energy states $\ket{0}$ and $\ket{3}$ are in resonance, and introduce a generalized persistent current
\be
I_P^{0n} = (I_{nn} - I_{00})/2 \label{IP0n},
\ee
which captures state dependence of the current matrix element $I_{nn}$. As pointed out before, the broadening widths due to both low and high frequency noise are functions of the persistent current: $W \propto I^{0n}_P,\ \gamma \propto (I_P^{0n})^2$. One should therefore rescale these parameters in $G_L(\omega)$ and $G_H(\omega)$ according to \eqref{IP0n}. However, since the state dependence of the persistent current is expected to be weak, we assume $I_P^{0n} \approx I_P$ and neglect these small corrections.

We obtain $G_R(\omega)$ by calculating the transition rate when the only broadening effect is due to intrawell relaxation, i.e., other noise contributions are turned off ($W=\gamma=0$). To simplify the calculation we focus on a three-state system described by
\be
H_S = \sum_{n=0,1,3} E_n \ket{n}\bra{n} - {\Delta_{03} \over 2} (\ket{0}\bra{3} + \ket{3}\bra{0}) \label{H3level}
\ee
with interaction Hamiltonian
\be
H_{\rm int} = -Q_R \ket{3}\bra{1}  + h.c.
\ee
where 
\be \label{QR}
Q_R = I_{31} \delta \Phi + V_{31} \delta q
\ee
provides coupling to flux and charge noise. Notice that the interaction Hamiltonian can only cause transition between states $\ket{3}$ and $\ket{1}$. The intrawell relaxation rate can be calculated using Bloch-Redfield theory
\be
\Gamma_{31}(\omega_{31}) = S_R(\omega_{31}),
\ee
where
\ba
S_R(\omega) =  \int dt \, e^{i\omega t}\langle Q_R(t)Q_R(0)\rangle
\label{SR}
\ea
is the environment spectral density corresponding to $Q_R$ defined in \eqref{QR}. 
When this relaxation is strong, it is not possible to separate interwell tunneling and intrawell relaxation as two independent processes. We therefore combine them into a single quantum mechanical process that creates transition from $\ket{0}$ to $\ket{1}$ mediated by state $\ket{3}$ (via virtual transition). 

Using perturbation expansion in $\Delta_{03}/\omega_{03} \ll 1$, we diagonalize Hamiltonian \eqref{H3level} to obtain
\be
H_S = \sum_{n=0,1,3} \tilde{E}_n \ket{\tilde n}\bra{\tilde n},
\label{HSpert}
\ee
where $\tilde{E}_n$ are perturbed energies and
\ba
\ket{\tilde 0} \!\!&=&\!\! \ket{0} + {\Delta_{03} \over 2\omega_{03}} \ket{3}, \\
\ket{\tilde 1} \!\!&=&\!\! \ket{1}, \\
\ket{\tilde 3} \!\!&=&\!\! \ket{3} - {\Delta_{03} \over 2\omega_{03}} \ket{0},
\ea
are perturbed eigenstates. The interaction Hamiltonian in this basis is
\ba
H_{\rm int} =  -Q_R \left(  \ket{\tilde 3}\bra{\tilde 1} + {\Delta_{03} \over 2\omega_{03}} \ket{\tilde 0}\bra{\tilde 1} \right) + h.c.
\label{Hintpert}
\ea
Since there is no off-diagonal term between $\ket{\tilde 0}$ and $\ket{\tilde 3}$ in Hamiltonians \eqref{HSpert} and \eqref{Hintpert}, we can now remove state $\ket{\tilde 3}$ from consideration. Introducing Pauli matrices
\ba
\sigma_x &=& \ket{\tilde 0}\bra{\tilde 1} + \ket{\tilde 1}\bra{\tilde 0}, \nn
\sigma_z &=& \ket{\tilde 1}\bra{\tilde 1} - \ket{\tilde 0}\bra{\tilde 0},
\ea
we obtain the familiar two-state Hamiltonian
\ba
H = -{1\over 2} \epsilon\, \sigma^z - {\Delta_{03} \over 2\omega_{03}}  Q_R \, \sigma^x,
\ea
where $\epsilon = \omega_{01} = \omega_{03} {+} \omega_{31}$ is the energy bias from qubit degeneracy. Here, we ignore second order corrections to energies $\tilde{E}_n$. The $\sigma^x$ term can be treated using Bloch-Redfield formalism to obtain
\be
\Gamma_{03}(\omega_{03}) = {\Delta_{03}^2 \over 4\omega_{03}^2} S_R(\omega_{03} {+} \omega_{31}).
\label{Gammaintrawell}
\ee
Equation \eqref{Gammaintrawell} is divergent at $\omega_{03} = 0$. We can remove the divergence the same way as in \eqref{GH0} by writing
\be
\Gamma_{03}(\omega_{03}) = {\Delta_{03}^2 \over 4} G_R(\omega_{03}),
\ee
with the envelope function
\be
G_R(\omega) 
= {\Gamma_{31}(\omega {+} \omega_{31}) \over \omega^2 + [\Gamma_{31} (\omega {+} \omega_{31})/2]^2}, %
\label{GR}
\ee
where $\Gamma_{31}(\omega) = S_R(\omega)$ is a frequency dependent intrawell relaxation rate. As we mentioned before, intrawell relaxation can be caused by both flux and charge noise. Therefore, $S_R(\omega)$ could be a sum of two components
\be
S_R(\omega) = I_{31}^2 S_\Phi(\omega) + V_{31}^2 S_q(\omega),
\ee
where $I_{n1}$ and $V_{n1}$ are defined in \eqref{IVnm}. However, with our noise parameters, contribution of flux noise is negligible compared to charge noise. We use \eqref{Sqsgn} for spectral density of charge noise, with magnitude of noise characterized by a constant loss tangent $\tan \delta_C$. To avoid singularity at $\omega = 0$, we need to replace ${\rm sgn}(\omega)$ with a smoother function. The center of the MRT peaks is dominantly broadened by the low-frequency flux noise, with almost no effect from low-frequency components of charge noise. We therefore choose 
\be
{\rm sgn}(\omega) \to \tanh(\omega/T)
\ee
which gives maximally flat spectrum $S_q(\omega)$ near $\omega = 0$ without additional fitting parameters. We therefore write
\be
\Gamma_{31}(\omega) = \zeta \, { \tanh (\omega/T) \over 1 - e^{-\omega/T}}. \label{SRomega}
\ee
where 
\be \label{IWME}
\zeta  = 2C V_{31}^2 \tan \delta_C.
\ee
is now a frequency independent parameter characterizing the width of $G_R(\omega)$ in \eqref{GR}.  $\zeta$ measures the (frequency independent) intrawell relaxation out of state $\ket{3}$. If we ignore frequency dependence of \eqref{SRomega} we recover the symmetric result of [\!\!\citenum{amin2008macroscopic}]. The relaxation rate then would not satisfy detailed balance and cannot explain the experimental results.

For higher energy states in the target well, there are more than one channel of relaxation. Therefore, the effective broadening becomes larger with higher energy. A more general and rigorous derivation of \eqref{GR} is provided in [\!\!\citenum{theorypaper}]. 

\section{Simplified model}
\label{AppendixD}

\begin{figure}
\begin{center}
\includegraphics[width=\columnwidth]{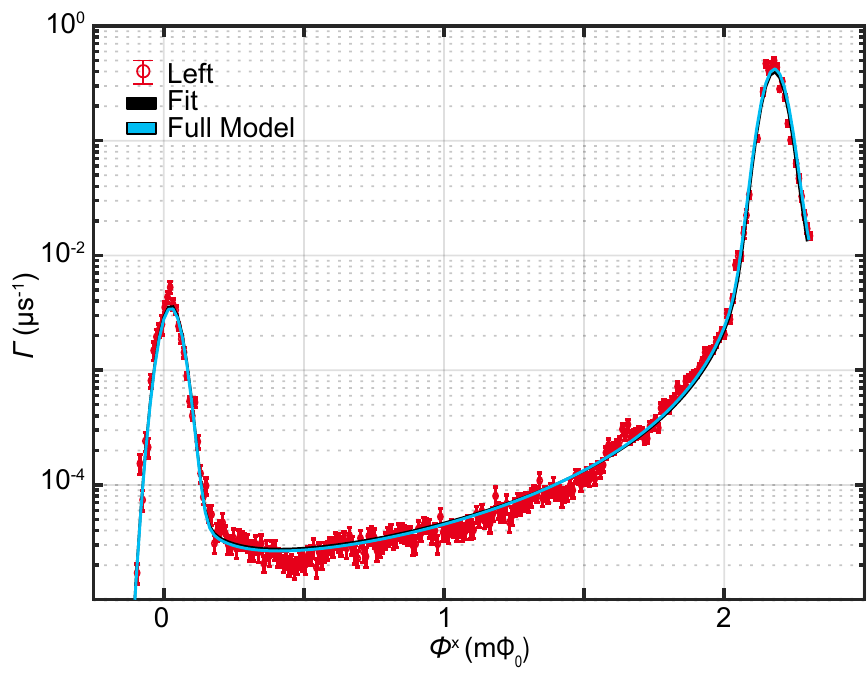}
\caption{\label{fig5} Comparison between the simplified model (black) used to fit the experimental data (red symbols) and the full model (blue). The simplified model uses approximations (detailed in appendix D) to reduce computational time. These approximations are relaxed by using the full model (see main text) that uses the extracted noise parameters from the fit. The two models agree very well with each other and with the experimental data.}
\end{center}
\end{figure}

The formalism described in the previous sections was obtained using Hamiltonian \eqref{h32} and interaction Hamiltonian \eqref{hint2}. These Hamiltonians were themselves obtained from the rf-SQUID Hamiltonian \eqref{HS} after partial diagonalization. The procedure to extract Hamiltonian parameters is time consuming and requires accurate knowledge of circuit parameters, such as inductance, capacitance, and critical current. In this section we introduce an approximation to this model that allows fitting to experimental data and extracting noise parameters without diagonalization or knowledge of rf-SQUID parameters. The assumptions behind this approximation are as follows:
\begin{enumerate}

\item Energy bias, $\epsilon = E_0 - E_1$, is a linear function of the applied flux $\efb$ (as in \eqref{ep7}) over the experimental range.

\item Persistent current $I_P$ has weak bias dependence and is measured independently.

\item Tunneling amplitudes $\Delta_{0n}$ have negligible bias dependence.

\item Current matrix element $I_{nn}$ ($\approx I_P$) is weakly dependent on state $\ket{n}$, therefore, noise parameters $W$ and $\gamma$ are the same for all $\Gamma_{0n}(\epsilon)$.

\item Capacitive loss tangent $\tan \delta_C$ is constant over the range of frequencies that matter for intrawell relaxation (close to $\omega_{31}$).

\item Inter- (intra-) well transitions are dominantly affected by flux (charge) noise.

\end{enumerate}

With these assumptions, one can fit the model to the experimental data using six fitting parameters, $\Delta_{01}$, $\Delta_{03}$, $\omega_{31}$, $W$, $\gamma$, and  $\zeta$, with no need for diagonalization. Note that all these parameters are in energy units. However, the potential tilt is applied to the rf-SQUID via an external flux bias ($\epsilon = 2I_P\Phi^x$). Therefore, the MRT peaks are measured as functions of flux (not energy) bias. The distance between the MRT peaks, $\Phi^x_{31} = {\omega_{31}/ 2I_P}$, is also directly measured in flux units. It is therefore convenient to express noise parameters directly in flux units:
\be
W_\Phi = {W \over 2I_P}, \qquad 
\gamma_\Phi = {\gamma \over 2I_P}, \qquad
\zeta_{\Phi} = {\zeta \over 2 I_P}.
\ee
Each of these broadening parameters characterizes one component of noise. $W_\Phi$ measures the r.m.s. value of the low-frequency flux noise, and  $\gamma_\Phi$ measures the magnitude of the high frequency flux noise. From $\gamma_\Phi$, the dimensionless ohmic coefficient and the inductive loss tangent can be calculated:
\be
\eta = {4I_P \gamma_\Phi \over T}, \qquad 
\tan \delta_L(\omega) = {\gamma_\Phi \over 4L I_P} \left|{\omega \over T}\right|.
\ee
Finally, the broadening parameter $\zeta_{\Phi}$ characterizes the charge noise. When the potential barrier is high, the bottom of the target well can be approximated by a parabola. The lowest energy levels inside the well can therefore be obtained using a harmonic oscillator model. One can then show that
\be
V_{31} \approx \sqrt{\omega_{31} \over 2C}.
\ee
Using \eqref{SRomega} and \eqref{IWME}, and assuming $\omega_{31} \gg T$, we obtain 
\be
\Gamma_{31}(\omega_{31}) \approx \zeta \approx \omega_{31} \tan \delta_C.
\ee
This is what one expects for relaxation in a harmonic oscillator. Converting to flux units, we obtain
\be
\tan \delta_C \approx {\zeta_\Phi \over \Phi^x_{31}}.
\ee
As usual, loss tangent is the ratio of peak broadening and oscillation frequency, both measured in flux units.

\bibliography{MRT-Manuscript-bib}{}

 \end{document}